\documentclass[letterpaper,12pt, reqno, utf8,latin9]{article}
\usepackage{sourceserifpro}
\usepackage[T1]{fontenc}
\usepackage{amsmath,amssymb,amsthm,eucal,bbold,bm}
\usepackage[italic,eulergreek]{mathastext}

\usepackage[hyphens]{url}
\MTsetmathskips{f}{\thinmuskip}{0mu}
\MTsetmathskips{y}{\thinmuskip}{0mu}
\MTsetmathskips{p}{\thinmuskip}{0mu}
\MTsetmathskips{l}{0mu}{\thinmuskip}
\MTsetmathskips{j}{\thinmuskip}{\thinmuskip}

\usepackage{titlesec}
\titleformat*{\section}{\centering\large\bfseries}
\titleformat*{\subsection}{\bfseries}
\titleformat{\paragraph}[runin]{\itshape}{}{}{}[]
\titlelabel{\thetitle.\quad}

\makeatletter
\renewcommand{\paragraph}{%
  \@startsection{paragraph}{4}%
  {\z@}{2.1ex \@plus 1ex \@minus .2ex}{-1em}%
  {\normalfont\normalsize}%
}
\makeatother

\usepackage[onehalfspacing]{setspace}
\AtBeginDocument{\frenchspacing}
\usepackage{microtype}
\usepackage[left=1in, right=1in, top=0.5in, bottom=1in]{geometry}

\usepackage{enumitem}			
\usepackage{lscape}
\usepackage[title,titletoc 
			]{appendix}
\usepackage[stable,bottom]{footmisc}	

\usepackage{advdate}

\usepackage{ccicons}			

\usepackage{multicol}			

\usepackage{amsmath, amsfonts} 	
\usepackage{amsthm,amssymb}
\usepackage{bbm}				
\usepackage{siunitx}			
\usepackage{textcomp}			
\usepackage{dsfont}				
\usepackage{bm}

\usepackage{graphicx}
\usepackage{caption}
\usepackage{subcaption}
\usepackage{hyperref}			
\usepackage{multirow}			
\usepackage{booktabs}			
\usepackage{arydshln}			

\usepackage{placeins}			

\usepackage{tikz}
\usetikzlibrary{decorations.pathmorphing}
\usetikzlibrary{shapes.geometric, arrows}
\usetikzlibrary{arrows,positioning,shapes,decorations.pathmorphing,snakes}
\tikzset{
  yn/.style={draw,thick,rounded corners,fill=yellow!20,inner sep=.3cm},
  bn/.style={draw,thick,rounded corners,fill=UBCblue!20,inner sep=.3cm},
  on/.style={draw,thick,rounded corners,fill=UBCblue!20,inner sep=.3cm},
  rn/.style={draw,thick,rounded corners,fill=UBCblue!20,inner sep=.3cm},
  greenn/.style={draw,thick,rounded corners,fill=green!20,inner sep=.3cm},
  grayn/.style={draw,thick,rounded corners,fill=gray!20,inner sep=.3cm},
  to/.style={
    ->,>=stealth',shorten >=1pt,semithick,font=\sffamily\footnotesize
  },
  from/.style={
    <-,>=stealth',shorten >=1pt,semithick,font=\sffamily\footnotesize
  },
  tofrom/.style={
    <->,>=stealth',shorten >=1pt,semithick,font=\sffamily\footnotesize
  },
  every node/.style={align=center},
  squig/.style={->,line join=round,decorate, decoration={zigzag,
    segment length=8,amplitude=2,post=lineto,post length=2pt}}
}
\usepackage{apacite}

\newcommand\YUGE{\fontsize{30}{40}\selectfont}

\usepackage{tabularx,booktabs,siunitx}
\newcolumntype{C}{>{\centering\arraybackslash}X} 
\usepackage[flushleft]{threeparttable}

\vspace{-1cm}

\title{\YUGE  \textbf{Does personality affect the allocation of resources within households? }}

\vspace{-1cm}
\author{
\textit{[Version: \today]}  
}
\date{Gastón P. Fernández\footnote{Ph.D. student at the University of Leuven (KU Leuven), Department of Economics, Naamsestraat 69, box 3565, 3000 Leuven (e-mail: gfernandez@kuleuven.be). I deeply appreciate the invaluable guidance of my advisors Laurens Cherchye and Frederic Vermeulen. I would also like to thank Wietse Leleu and all participants at the Conference of the European Society for Population Economics (ESPE) in Belgrade, the Trans-Atlantic Doctoral Conference (TADC) in London, and the Public-Labor-Health Seminar, the Household Economics Gathering, and the ECORES Summer School in Leuven for their helpful comments. All errors are on my own.} \\ 
University of Leuven (KU Leuven)}

\begin{document}
\maketitle

\vspace{-6mm}
\abstract{This paper examines whether personality influences the allocation of resources within households. To do so, I model households as couples who make Pareto-efficient allocations and divide resources according to a distribution function. Using a sample of Dutch couples from the LISS survey with detailed information on consumption, labor supply, and personality traits at the individual level, I find that personality affects intrahousehold allocations through two channels. Firstly, the \textit{level} of these traits act as preference factors that shape individual tastes for consumed goods and leisure time. Secondly, by testing distribution factor proportionality and the exclusion restriction of a conditional demand system, I observe that \textit{differences} in personality between spouses act as distribution factors. Specifically, these differences in personality impact the allocation of resources by affecting the bargaining process within households. For example, women who are relatively more conscientious, have higher self-esteem, and engage more cognitively than their male partners receive a larger share of intrafamily resources.

\vfill

}

\newpage
\section{Introduction}


There is increasing evidence that personality traits matter for relevant life outcomes \cite{heckman2021some}. For instance, personality is associated with the formation of future cognitive skills \cite{cunha2010estimating} or with the educational and occupational choices over the life cycle \cite{todd2020dynamic}. Personality is also correlated with the probability of marriage and divorce \cite{lundberg2012personality} and is a relevant attribute on which individuals sort into the marriage market \cite{dupuy2014personality}. Nevertheless, much less is currently known about personality's impact on intrahousehold consumption patterns. For example, do personality traits affect the allocation of resources through their impact on individual preferences over goods? Or are there other mechanisms by which personality might shape the way couples decide over total resources? Is personality related to the distribution of power within households? 

In this paper, I aim to empirically investigate the questions mentioned above by structurally testing the role of personality traits in resource allocation within households. I show that \textit{differences} in personality traits between spouses play a significant role in shaping the distribution of resources within established households. Families are modeled as couples who make static decisions regarding private and public consumption and also allocate their time to the labor market. As a starting point, I assume that each adult household member has their own rational preferences. Additionally, I assume that couples make Pareto-efficient allocations and distribute resources among household members through an intrahousehold decision process \cite{chiappori1988rational, chiappori1992collective}. By adopting this framework, I can test the concept of collective rationality, which refers to the collective model, using observed household allocations. This approach allows me to uncover relevant information underlying the consumption process. The main focus of this paper is to explore the hypothesis that personality traits may partially determine how couples divide resources. To investigate this, I test various theoretical restrictions of the collective model as formalized by \citeA{bourguignon2009efficient}. The collective framework not only enables the characterization of couples in terms of rational decisions but also allows for the integration of individual personality into a model of household consumption and labor supply.

 This article contributes theory-based evidence about new channels that may explain consumption inequality within households. In the collective model, couples maximize a weighted sum of individual utilities, where the weights are referred to as Pareto weights. In the collective literature, when examining the impact of a specific variable on household behavior, a distinction is made between two channels: \textit{preference} and \textit{distribution} factors. Preference factors typically influence individual preferences for consumed commodities, while distribution factors specifically affect the decision-making process within the household through changes in the Pareto weights.  In this sense, the \textit{level} of a specific variable (e.g., years of schooling) is often considered as a preference factor and the \textit{relative} amount of it (e.g., differences in education between partners) as a distribution factor. I leverage this notion and investigate both distribution factor proportionality and the exclusion restriction of a conditional demand system by utilizing differences in personality between spouses. I test the testable restrictions derived from collectively rational behavior and find no evidence to reject that differences in personality influence the bargaining process. Furthermore, I demonstrate that certain personality factors, such as differences in conscientiousness or self-esteem between spouses, are strongly associated with consumption inequality within the household. These findings provide valuable insights into the role of personality traits in shaping intrahousehold resource allocation dynamics.


 Distribution factors, which influence household decisions without directly impacting preferences, have been extensively studied in the collective literature. These factors encompass a wide range of variables, including relative wages among spouses and the presence of divorce laws in relevant matching markets. For instance, \citeA{browning1994income} demonstrate that the intrahousehold allocation of resources is related to factors such as relative ages and relative incomes in consumption models.  \citeA{chiappori2002marriage} extend earlier versions of the collective model and test their implications by introducing the local sex ratio and divorce laws as distribution factors in a labor supply model.  In a nonparametric setting, \citeA{cherchye2011revealed} examine the relationship between the intrahousehold share of income and differences in age and educational level between spouses. Furthermore, exploiting exogenous variation from a randomized cash transfer program in Mexico, several studies have constructed distribution factors and tested the theoretical restrictions of the collective model (see \citeA{bobonis2009allocation, attanasio2014efficient, de2022household}).\footnote{See \citeA{browning2014economics} for a comprehensive review. }

Building upon the collective framework and the existing applied research on the impact of personality, this paper contributes novel evidence suggesting that both intrahousehold rational behavior and consumption inequality are linked to the personality types of household members. While recent advancements in personality research have been extensively reviewed (see \citeA{john2010handbook}), the detailed examination of this particular issue is still relatively unexplored. In a related study, \citeA{flinn2018personality} develop a model of household behavior and apply it to Australian data to investigate how personality traits influence cooperative and non-cooperative interactions within households, as well as members' labor supply and wage rates. Their findings demonstrate that personality directly affects intrahousehold behavior and also indirectly impacts individual wages. The approach taken in the present paper differs from \citeA{flinn2018personality}. Instead of applying a behavioral model to the data, this study leverages a set of testable restrictions derived from \citeA{bourguignon2009efficient}, which serve as necessary and sufficient conditions for the collective model. By adopting this approach, the current study can structurally test the extent to which personality traits determine the allocation of resources between partners by influencing their respective bargaining positions within the household.

The rest of the paper unfolds as follows. Section 2 provides an introduction to the notation used and presents a collective model of household consumption and labor supply. This section also outlines the testable restrictions of the model based on observed household behavior, specifically focusing on distribution factor proportionality and the exclusion restriction of a conditional demand system. In Section 3, the sample used in the analysis is described, along with the available measures of personality traits. Section 4 outlines the empirical strategy employed in the study. It discusses how potential issues of multicollinearity in personality traits are addressed. Furthermore, it presents the functional form for the household demand functions and explains how tests of the collective model are derived from these functions. Section 5 presents the results obtained from testing the restrictions of the collective model. In Section 6, the relationship between intrahousehold consumption inequality and personality traits is discussed. Finally, Section 7 concludes the paper.

\section{Theory}
The analysis considers households consisting of two adult members: the wife ($f$) and the husband ($m$). These individuals jointly make consumption decisions involving a Hicksian public good ($C \in \mathbb{R}_+$), private Hicksian assignable goods for each member ($c^i \in \mathbb{R}_+$), and individual leisure time ($\ell^i = T - L^i$), where $\ell^i \in \mathbb{R}_+$ represents the amount of leisure time, $T$ is the time endowment for each individual, and $L$ is the time supplied to labor ($i=m, f$). It is assumed that children, if present, do not participate in the allocation of the household budget. The prices of all Hicksian goods are normalized to one and wages ($w^i \in \mathbb{R}_{++}$) represent the prices of individual leisure.
The preferences of household members are captured by well-behaved utility functions. Each individual has an egoistic utility function denoted as $u^i(c^i, \ell^i, C; \bm{\xi})$. The utility function also depends on the vector $\bm{\xi}$, which represents observed heterogeneity (i.e., taste shifters).


\pagebreak
In the collective model of \citeA{chiappori1988rational, chiappori1992collective}, any Pareto-efficient intrahousehold allocation can be characterized as the solution of the following optimization program:
 \begin{equation}\tag{P1}
 \begin{split}
\max_{c^m,c^f, \ell^m,\ell^f,C} &  \biggr[ u^m(c^m,\ell^m, C;\bm{\xi})  
 + \mu(w^m,w^f, y,\textbf{z}) u^f(c^f,\ell^f, C;\bm{\xi})\biggr] \\ 
	  \textrm{s.t.}& \quad c^m+c^f+C+w^m\ell^m+w^f\ell^f \leq y, \\
	&  \quad \hspace{4.2cm}  c^i \geq 0, \\
	&  \quad \hspace{4.2cm}C \geq 0, \\
	&  \quad \hspace{3.4cm}T \geq \ell^i \geq 0, \\
\end{split}
\end{equation}
where $y$ is household full income defined by $y= w^mT+w^fT + x$ with $x\in {\rm I\!R}_+$ the household nonlabor income, and $\mu \in$  $]0, 1[$ in the objective function is the Pareto weight that depends on (exogenous) wages, income, and distribution factors ($\textbf{z}$). A variation on elements of $\textbf{z}$ could impact outside options of household members and thus their intrahousehold bargaining power (see \citeA{vermeulen2002collective}).\footnote{In axiomatic bargaining models, variables that are only applicable for threat points of the bargaining process can be potential distribution factors. See the discussion about extrahousehold environmetal parameters in \citeA{mcelroy1990empirical} and about bargaining models in \citeA{browning2014economics}.} I take both household composition and intrafamily allocation of power as exogenously given. The solution to (P1) implies a set of differentiable household demand functions for goods and leisure that depend on prices, full income, observed heterogeneity, and the distribution function:

	\begin{equation}
\textbf{g}=\textbf{g}\Bigr[w^m, w^f, y,\mu(w^m, w^f,y,\textbf{z});\bm{\xi}\Bigr] \quad\quad \forall \hspace{2mm} \textbf{g} \in \{ \textbf{c}, \bm{\ell}, C\}.
	\end{equation} 	\vspace{0mm}

\textbf{Distribution factor proportionality.} As explained by \citeA{bourguignon2009efficient}, in a setting with no price variation distribution factor proportionality is necessary and sufficient for the collective model.\footnote{The first notions of the proportionality condition with only private consumption are introduced in \citeA{bourguignon1993intra} and \citeA{browning1994income}.  \citeA{bourguignon2009efficient} extend these results for public goods and externalities in consumption.} This entails testing a set of cross-equation restrictions based on the estimation of the household demand system (1):


\vspace{2mm}
\noindent

	\begin{equation}
		\frac{\partial c^m/\partial z_1}{\partial c^m/\partial z_k}=		\frac{\partial c^f/\partial z_1}{\partial c^f/\partial z_k}=		\frac{\partial \ell^m/\partial z_1}{\partial \ell^m/\partial z_k} =
		\frac{\partial \ell^f/\partial z_1}{\partial \ell^f/\partial z_k} =
		\frac{\partial C/\partial z_1}{\partial C/\partial z_k} \quad\quad \forall \hspace{2mm} k = 2, \dots, K.
	\end{equation} \vspace{2mm}

The intuition of equation (2) is that distribution factors ($\textbf{z}$) only affect the intrahousehold allocation of consumption and leisure through their impact on the distribution function ($\mu$). To see this, take the marginal change in distribution factor $z_k$ on the household demand for commodity $j$:
\vspace{2mm}
\begin{equation}
	\begin{split}
		\frac{\partial g_j}{\partial z_k} & = \frac{\partial g_j}{\partial \mu} \frac{\partial \mu}{\partial z_k}.
	\end{split}
\end{equation} \vspace{0mm}

Comparing the effect of two distribution factors, $z_k$ and $z_l$, we get:

\begin{equation}
	\begin{split}
		\frac{\partial g_j/\partial z_k}{\partial g_j/\partial z_l}
		& = \frac{\partial \mu / \partial z_k}{\partial \mu / \partial z_l},
	\end{split}
\end{equation}

\vspace{2mm}
\noindent
where the right-hand-side term in equation (4) is independent of the demand for good $j$. 

\vspace{5mm}
$z$-\textbf{conditional demand system.} An alternative demand system is the $z$-conditional system coined by \citeA{bourguignon2009efficient}. Under the assumption that distribution factor $z_1$, say, is strictly monotonic on commodity $c^m$, say, it is possible to invert the demand function for such good on this (continuous) factor:

\begin{equation}
	z_1 = v(w^m, w^f,y, c^m,\textbf{z}_{-1};\bm{\xi}),
\end{equation}

\vspace{2mm}
\noindent
where $\textbf{z}_{-1}$ is equal to $\textbf{z}$ but excluding the first element.\footnote{Appendix B provides evidence that supports monotonicity between male private consumption and one of the distribution factors presented in Section 4.}  Substituting (5) into the demand for the remaining goods $\Phi(\cdot)$, we get the $z$-conditional demand system for  $\tilde{\textbf{g}}$ with $\tilde{\textbf{g}} \in \{ c^f, \bm{\ell}, C\}$:

\begin{equation}
\begin{split}
	\tilde{\textbf{g}} &= \Phi(w^m, w^f,y,\textbf{z};\bm{\xi}),\\
	&= \Phi\Bigr[w^m, w^f,y,v(w^m, w^f,y, c^m,\textbf{z}_{-1};\bm{\xi}),\textbf{z}_{-1};\bm{\xi}\Bigr],\\
	&= \tilde{\textbf{g}}(w^m, w^f,y,c^m,\textbf{z}_{-1};\bm{\xi}).\\ 
\end{split}
\end{equation}

The restriction of the collective model based on the estimation of the (conditional) demand system in equation (6) states that subject to the conditioning good ($c^m$), the demand for the remaining goods should be independent of all other distribution factors. This translates into the following testable implication:
\begin{equation}
	\frac{\partial \tilde{\textbf{g}}(w^m, w^f,y,c^m,\textbf{z}_{-1};\bm{\xi})}{\partial z_k} = 0 \quad\quad \forall \hspace{2mm} k = 2, \dots, K.
\end{equation}
The restriction described in equation (7) implies that, conditional on the commodity used to invert $z_1$, additional distribution factors should not provide any meaningful additional information about the intrahousehold behavior. It is important to note that for this restriction to have empirical significance, it requires at least two distribution factors and at least two demand functions.

Although the testable implication in equation (7) is empirically more powerful than implication (2), which is used as a robustness check in the empirical application, both restrictions capture the same underlying mechanism.\footnote{See Proposition 2 in \citeA{bourguignon2009efficient} and the discussion thereof.} The intuition behind this is illustrated in Figure 1. Suppose we observe an optimal household demand function that is relatively more representative of $m$'s preferences, such as $\textbf{g}^0$. Now, assume that we want to reallocate intrahousehold resources in a manner that is more favorable to the wife's ($f$) preferences, resulting in household decisions represented by $\textbf{g}^1$. The testable restrictions of the collective model inform us that variations in the distribution factors $\textbf{z}$ would only impact such a reallocation of resources by shifting the individual bargaining weights ($\mu$). In other words, distribution factors do not alter the Pareto frontier since they do not directly affect preferences or the budget constraint.

\begin{figure}[hbt]
\caption{The collective effect}
\centering

\begin{tikzpicture}[scale=1]

\draw [<->] (0,5) node [above] {$u^m$} -- (0,0) -- (5,0) node [right] {$u^f$};
\draw (0,4.1) to [out=-15, in=120] (2.8,2) to [out=-60, in=95] (3.4,0);

\draw [->] (3.4,0.2) -- (4.5,0.4);
\node[right] at (4.4,0.5) {\scriptsize{Utility Pareto frontier}};


\node [right] at (0.6,4.2) {\footnotesize{ $\textbf{g}^0 = (\tilde{\textbf{c}},\tilde{\bm{\ell}}, \tilde{C})$}};
\draw [dashed] (0.2,4.15) -- (2.6,3);
\draw [fill] (1,3.75) circle [radius=.06];

\node [right] at (2.8,2) {\footnotesize{$\textbf{g}^1= (\hat{\textbf{c}},\hat{\bm{\ell}}, \hat{C})$}};
\draw [dashed] (2.45,2.7) -- (3.6,0.5);
\draw [fill] (2.8,2) circle [radius=.06];

\draw [->] (4,4) to [bend left = 20] (5,2.6);
\node [text width=4.5cm, right] at (3,3.5) {$\frac{\partial \mu(\textbf{w}, y, \textbf{z})}{\partial z_k}$};

\end{tikzpicture}
\parbox{\textwidth}{\footnotesize%
\vspace{1mm} 
\centering
\scriptsize{Source: Based on \citeA{browning2014economics}.}}
\end{figure}
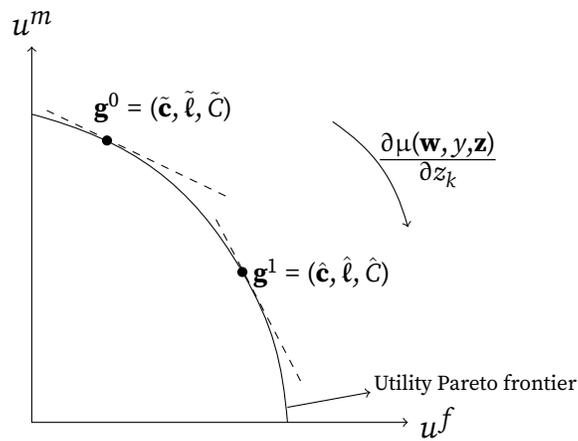


\section{Data}
I use a sample of Dutch households obtained from the Dutch Longitudinal Internet Studies for the Social sciences (LISS) panel gathered by CentERdata. This dataset provides rich information on economic and sociodemographic variables. Crucially, it also collects detailed data on individual consumption and a set of member-specific personality scales.

The sample selection criteria for this study are as follows, similar to those used in other studies such as \citeA{cherchye2017household} and \citeA{cherchye2012married}. Couples included in the sample must have both adults between the ages of 25 and 65. Both adults in the couple must participate in the labor market for at least 10 hours per week, as wage information is required. Couples with at least one self-employed adult are excluded from the sample. This is because obtaining wage information for self-employed individuals is more complex compared to salaried workers. The sample includes only couples with no additional household members apart from children residing in the household. For example, couples living with friends or parents are excluded. Due to significant imbalance issues in the panel structure of the data, the study does not make use of the panel structure and treats the data as a pooled cross-section. Overall, the sample consists of 1130 couples pooled from five different years, ranging from 2009 to 2015.

Table 1 provides summary statistics for the main variables used in the analysis. All economic variables are in weekly real terms. Full income is defined as the sum of spouses' wages multiplied by the total time available (i.e., 112) plus any non-labor income of the household. Leisure for each partner is derived by subtracting the hours worked by each individual from the total available time. The dataset includes information on assignable consumption for each household member. This refers to individual expenditures on various goods such as food, tobacco, or clothing. In the empirical analysis, these individual expenditures are treated as a Hicksian aggregate commodity. Total household private consumption represents the sum of both spouses' total private consumption, including their individual assignable consumption. Household consumption is calculated as the sum of public consumption and assignable private consumption. Public expenses, such as mortgage payments, are considered as a Hicksian aggregate commodity. As shown in Table 1, females work fewer hours and have lower wages compared to males. In terms of assignable consumption, females spend slightly more per week than males. The majority of total household consumption comes from public expenses. Females allocate more time to leisure activities than males, although a detailed breakdown of non-labor time is not available.\footnote{Data about the individual time allocated to household chores is only available in three waves.} Demographically, males are slightly older and have a higher educational level compared to females.

The spouses' personality traits in this study are measured using three different sources. The first source is Rosenberg's Self-Esteem Scale \cite{rosenberg1965society}, which assesses individuals' perceptions of their self-worth. The second source is the Need For Cognition Scale \cite{cacioppo1982need}, which serves as a proxy for an individual's inclination to engage in intellectual activities. The third source is the Big Five Personality Traits questionnaire \cite{goldberg1990alternative, goldberg1992development}, which captures personalities based on five overarching dimensions.\footnote{To construct each personality measure, I consider items with high loading values from exploratory factor analysis as in \citeA{flinn2018personality} and \citeA{todd2020dynamic}. These personality measures demonstrate high internal consistency, as indicated by Cronbach's alphas exceeding 0.7.} Out of the total 1130 couples in the sample, valid information on personality traits is available for 583 couples. For households with missing personality information, the values are imputed by averaging observed individual personality scores from other waves. This imputation approach takes into account the stability of personality traits over time, which has been suggested by previous studies.\footnote{See, e.g., \citeA{cobb2012stability}, \citeA{todd2020dynamic} or \citeA{fitzenberger2022personality}. See appendix A for the stability of personality traits in the current sample.} I test various imputation methods, such as using the median value, but the main results remain robust. Looking at the bottom of Table 1, on average, males tend to have higher values than females in measures of self-esteem, extraversion, and cognitive engagement. In contrast, females tend to score higher than males in conscientiousness, neuroticism, and agreeableness. Both males and females exhibit similar levels of openness. These gender differences in personality traits align with findings from previous studies conducted on Dutch samples (see, e.g., \citeA{nyhus2005effects} or \citeA{dupuy2014personality}). Importantly, the gender differences in personality traits observed in the sample remain virtually unchanged even after the imputation of missing personality traits.

\begin{table}[hbt!]
\caption{Summary statistics.}
\centering
\resizebox{0.7\textwidth}{!}{%
\begin{tabular}{lcccccccc} \hline \hline
				&\textbf{Mean}  &\textbf{Std. dev.}  & \textbf{Min} & \textbf{Max}  \\ \hline
\textit{\textbf{A. Economic variables:}} \\ \hline
\hspace{2mm}Male wage rate     & 13.63  &  3.71 &  6.88 & 29.90  \\
\hspace{2mm}Female wage rate   &  12.05 &  3.16 &  4.03 & 21.80  \\
\hspace{2mm}Male weekly hours worked  & 37.43 &  4.74  &12 & 60  \\
\hspace{2mm}Female weekly hours worked    & 25.98 &  7.99 &  10 & 48  \\
\hspace{2mm}Full income & 2820.69 &  576.79   & 1357.20 & 4770.11  \\
\hspace{2mm}Household private consumption  & 2241.59  &  472.04 &1142.50   & 4089.12  \\
\hspace{2mm}Assig. male private consumption  &   89.97 &51.78  &   15 &453.72  \\
\hspace{2mm}Assig. female private consumption   &   95.25 & 54.11  & 19.38   &507.66 \\
\hspace{2mm}Public consumption  &   579.10 &  229.75 &    102.96  & 1898.35 \\
\hspace{2mm}Total household consumption  &    764.32 &  256.07 &    173.21 &   2284.98 \\
\hspace{2mm}Male weekly  leisure  	 &  74.56 &   4.74 &   52  & 100  \\
\hspace{2mm}Female weekly leisure  	 &  86.01 &  7.99 &   64 & 102 \\\hline
\textit{\textbf{B. Demographic variables:}} \\\hline
\hspace{2mm}Male age 	  &  47.39&  9.76 &   25 & 65  \\
\hspace{2mm}Female age 	  &  45.46&  9.90 &   25 & 65  \\
\hspace{2mm}Number of children 	 			 &    1.16 & 1.11 &  0 & 5  \\
\hspace{2mm}Male dummy low education  		 &  .20 & .40 &  0 & 1  \\ 
\hspace{2mm}Female  dummy low education 	 &  .43 & .49 &  0 & 1  \\ 
\hspace{2mm}Male dummy middle education  	 &  .36 &.48 &  0 & 1  \\ 
\hspace{2mm}Female  dummy middle education 	 &  .23 & .42 &  0 & 1  \\ 
\hspace{2mm}Male dummy high education  		 &  .43 & .49 & 0 & 1  \\ 
\hspace{2mm}Female  dummy high education 	 &  .32 & .47 &  0 & 1  \\ \hline
\textit{\textbf{C. Personality traits:}} \\ \hline
\hspace{2mm}Male Openness &  3.07&    .26 &        1.37&        3.87\\
\hspace{2mm}Female Openness& 3.07&    .28 &          1.87&        3.87\\
\hspace{2mm}Male Extraversion& 3.18 &    .51 &        1.33 &        4.50 \\
\hspace{2mm}Female Extraversion&  3.12 &    .51 &        1.33 &        4.50\\
\hspace{2mm}Male Agreeableness & 3.07    &    .25 &        2.00 &        3.75 \\
\hspace{2mm}Female Agreeableness &  3.16 &    .20 &         2.37 &        3.75\\
\hspace{2mm}Male Neuroticism & 2.29&    .57 &        1.00 &           4.22 \\
\hspace{2mm}Female Neuroticism &  2.59 &    .59 &       1.00 &         4.33 \\
\hspace{2mm}Male Conscientiousness  & 2.78 &    .27 &        1.88 &        3.66 \\
\hspace{2mm}Female Conscientiousness  &2.85 &    .24 &           1.77 &        3.55\\
 \hspace{2mm}Male Self-esteem  &5.98 &    .65 &        3.80 &          7.00 \\
 \hspace{2mm}Female Self-esteem  &  5.85 &    .72 &        3.70 &          7.00\\
 \hspace{2mm}Male Cognitive engagement   & 4.78 &    .86&   2.66 &          7.00 \\
 \hspace{2mm}Female Cognitive engagement  &   4.39 &     .84 &          2.25 &   6.75\\
 
\hline \hline 

\end{tabular} }
\centering
  \parbox{\textwidth}{\scriptsize{
Notes: Sample size of 1130 couples. LISS waves 2009, 2010, 2012, 2015, and 2017 pooled up. All economic variables are in weekly 2015 euros.}}
\end{table}

\section{Empirical strategy}
In this section, I discuss the measures of \textit{relative} personality traits that are employed to examine the restrictions of the collective model outlined in Section 2. These relative personality traits capture the differences in personality between spouses within a household. The functional form for the household demand functions is also introduced. From these demand functions, several testable implications can be derived to assess the validity of the collective model. 

\vspace{3mm}
\textbf{Multicollinearity in personality traits.} To address the issue of multicollinearity arising from the seven measures of personality traits, I employ principal component (PC) analysis. This analysis is applied to the entire sample, which includes both women and men. The goal is to identify the principal components that explain the majority of the variance in the observed personality measures. By extracting these principal components, which are linearly uncorrelated factors, the paper aims to reduce the dimensionality of the personality traits and mitigate the multicollinearity problem. This approach allows for a more precise estimation of the effects of personality traits on intrahousehold consumption behavior \cite{jolliffe2002principal}. 

Table 2 presents the correlations between the principal components (PCs) and the individual personality measures, as well as the eigenvalues and the share of observed variance explained by each PC. The results indicate that the two principal components capture distinct aspects of personality traits. PC1 is associated with traits such as introversion, lower self-esteem, and lower cognitive engagement. On the other hand, PC2 is characterized by higher levels of neuroticism and conscientiousness. The eigenvalues and the proportion of observed variance explained by each PC reflect their relative importance in explaining the variability in the original personality measures.

For each couple in the sample, the relative endowment of personality traits between partners is calculated by constructing the ratio of spouses' principal components (PCs). These ratios, representing the relative distribution of personality traits, are treated as continuous measures and tested as distribution factors in the collective consumption model presented in Section 2. To facilitate comparison and analysis, the PCs are scaled from 1 to 100, considering that they can take negative values. Figure 2 displays the distribution of these ratios. On average, women tend to have higher values for the common personality factor represented by PC1 compared to men. In contrast, men exhibit higher values for PC2 relative to women.

\begin{table}[hbt!]
\caption*{Table 2. Principal components}
\centering
\begin{tabular}{lcc }

\hline \hline

\textbf{Personality}: & \textbf{PC1} & \textbf{PC2} \\
\hline
1. Extraversion 		 & $-$ & \\
2. Agreeableness  	 &  &\\
3. Openness    	 	 & & \\
4. Conscientiousness &  &$+$ \\
5. Neuroticism  		 & &$+$ \\
6. Self-esteem  		 & $-$  & \\
7. Cognitive engagement  		 &$-$  &\\
\hline 
Eigenvalue & 1.41 & 1.23 \\
Variance share & 28.58\% &21.75\%  \\
\hline \hline
\end{tabular}
\parbox{\textwidth}{\scriptsize%
\vspace{1eX} 
Notes: Explained share of the observed variance: 50.33\%. The table indicates the sign of those loadings that are larger than a cut-off of .8 with respect to the largest coefficient in each component (similar procedure as in \citeA{jolliffe2002principal}). The largest coefficient in PC1 is self-esteem; in PC2 is conscientiousness. }
\end{table}

\begin{figure}[hbt]
\caption*{Figure 2. Within-couple differences in personality traits}
\centering
\includegraphics[scale=.45,,trim={2mm 5mm 2mm 2mm},clip]{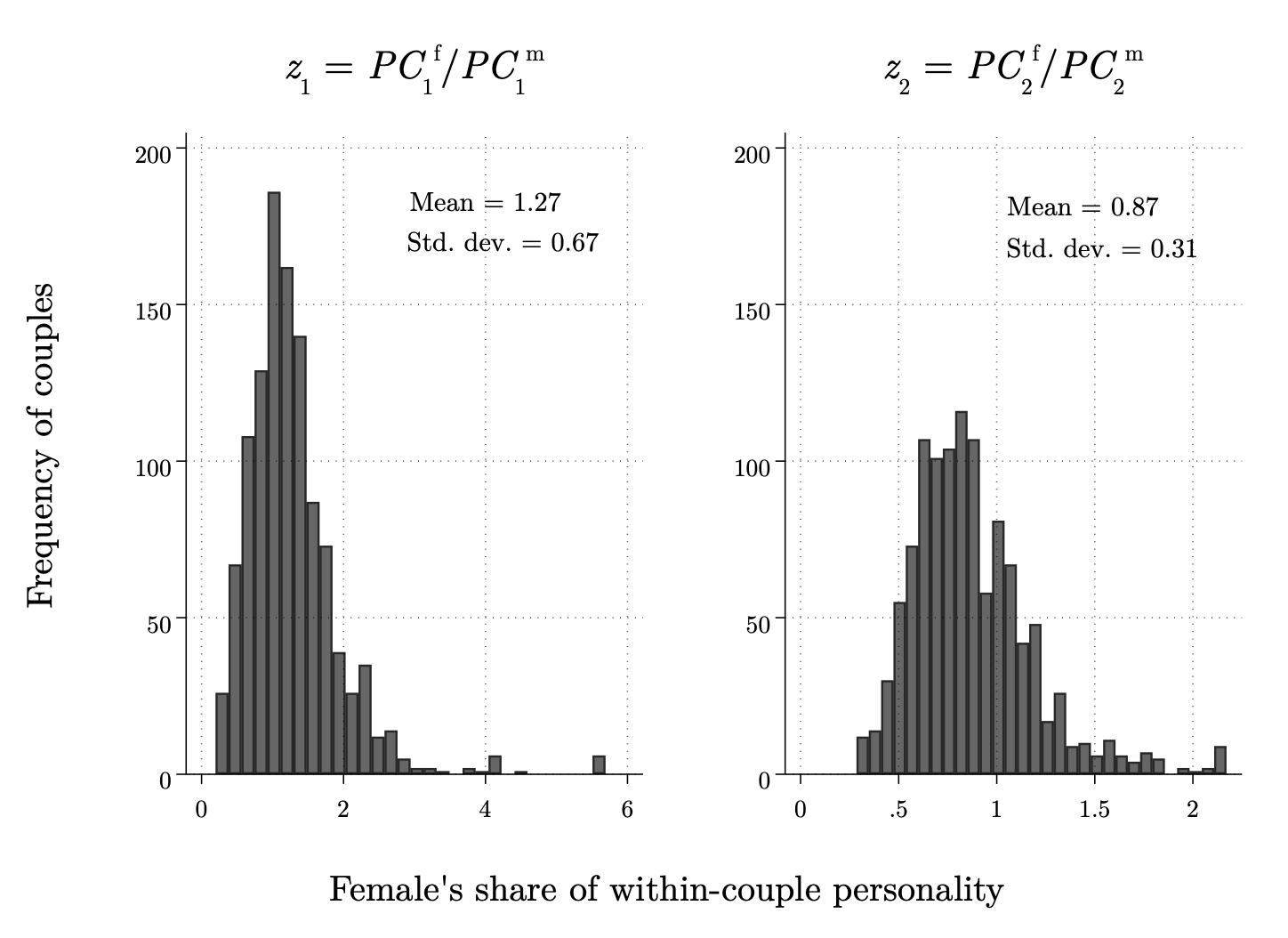}
\vspace{-1mm}
\end{figure}
\FloatBarrier

\vspace{3mm}
\textbf{Parametrization of unconditional demand functions.} To test the restrictions of the collective model, a functional form for the household demand functions needs to be specified. I follow \citeA{bobonis2009allocation} and parametrize the unconditional demand functions $\textbf{g} \in \{\textbf{c},\bm{\ell},C\}$ in budget share form as:
	\begin{equation}
	\omega_{j} = \alpha_{j} +\ln(\textbf{z}')\bm{\beta} + a_j(y) +b_j(y^2) + \ln(\textbf{w}')\bm{\lambda} +  \textbf{e}'\bm{\delta} +\textbf{m}'\bm{\psi} + \bm{\tau}_j + \varepsilon_{j} ,
	\end{equation} 
where for each couple in the sample, $\omega$ is the budget share on good $j$, $a$ and $b$ are functions of full income and its square, $\textbf{w}$ is a vector of partners' wages, $\bm{\tau}$ are time dummies capturing heterogeneity over time, and $\varepsilon$ is unobserved heterogeneity.\footnote{Potential sources of endogeneity for full income are measurement error in nonlabor income, taste shocks to total consumption that could be correlated to unobserved heterogeneity in the budget shares equations, or saving decisions that may be driving changes in nonlabor income.} Prices of composite goods, which are normalized to one, are assumed to enter through $\bm{\tau}$. The vector \textbf{z} includes the relative endowment of personality traits, i.e., ratios of PCs between partners of a household. The additional controls $\textbf{e}$ and $\textbf{m}$ are detailed below.\footnote{The assumption of a linear-log functional form allows for a straightforward interpretation of the coefficient estimates in the empirical model. Additionally, the empirical results remain consistent regardless of the specific functional form assumption chosen (results can be provided upon request).}

One potential source of endogeneity in equation (8) is the endogenous selection of couples in the marriage market, wherein individuals may form couples based on their respective personality traits. Despite the limitations of the current dataset, I address this potential issue in two ways.\footnote{Fully addressing selection in personality traits, such as through the estimation of a structural matching model, is beyond the scope of this paper.} First, the vector of taste shifters ($\textbf{e}$) includes, among other explanatory variables, the logarithm of the \textit{level} of principal components (PCs) of each spouse and their squares. I include the squares of the PCs to accommodate for potential nonlinearity in the influence of personality on preferences over commodities, as suggested in the analysis of \citeA{borghans2008economics}. Second, in all specifications, I incorporate the vector $\textbf{m}$ to account for marriage market conditions with respect to personality, as discussed in \citeA{dupuy2014personality}. This vector incorporates the weighted ratios of the number of husbands and wives who are of similar age and educational level and who have the same score in a given personality trait as the husband or wife of each household, divided by the corresponding number of husbands or wives. These ratios, referred to as \textit{personality ratios}, are akin to the sex ratio concept in \citeA{chiappori2002marriage} and serve to control for the underlying structure of the marriage market in the sample with respect to personality traits.

The proportionality restriction imposed by collective rationality (as expressed in equation (2)) on the system of unconditional demand functions can be formulated as follows:
	\begin{equation}
	\begin{split}
		\frac{\partial \omega_j/\partial \ln(z_{1})}{\partial \omega_j/\partial \ln(z_{2})} =& \frac{\partial \omega_s/\partial \ln(z_{1})}{\partial \omega_s/\partial \ln(z_{2})},  \\
		\frac{\beta_{j1}}{\beta_{j2}}=&\frac{\beta_{s1}}{\beta_{s2}} 
	\end{split}
	\end{equation}

\noindent 	
for all goods $j$, $s$, with $j \neq s$. If condition (9) is satisfied, it implies that there is no evidence to reject the hypothesis that the effects of \textit{differences} in personality traits between partners on resource allocation occur solely through their influence on the household's distribution function.

To test the nonlinear cross-equation restrictions presented in equation (9), the model is estimated as a system, allowing for correlation between the error terms across the budget shares equations. The cross-equation hypotheses are then examined using Wald test formulations. It is important to note that these formulations may be subject to statistical issues. For instance, in OLS systems, Wald tests tend to over reject the null hypothesis, and they are not invariant to the definition of the null hypothesis (see \citeA{greene2003econometric}). To address these concerns, this study adopts a similar approach to that of \citeA{bobonis2009allocation}. Firstly, the Wald tests are conducted using the bootstrap distribution with 1000 replications. Secondly, as a robustness check of the main results, linear Wald tests are computed based on the estimation of the $z$-conditional demand system proposed by \citeA{bourguignon2009efficient}.

\vspace{4mm}
\textbf{Parametrization of the $z$-conditional demand system.} Under the additional assumption that one distribution factor is strictly monotone in one good, we can derive the demand for that good as a function of the distribution factor. In my analysis, I find suggestive evidence indicating the presence of a monotonic correlation between factor $z_2 = PC_2^f/PC_2^m$ and male private consumption ($c^m$).\footnote{Refer to appendix B for detailed evidence on the monotonicity assumption. It is important to note that for the collective test based on the conditional demand system presented in this section, $z_2$ needs to be both continuous and statistically significant. For further discussion on this topic, see \citeA{de2022household}.} 

In budget share form, the demand for male private consumption ($c^m$) inverted on $z_2$ is given by:
	\begin{equation}
	\begin{split}
	\ln(z_{2}) =& \frac{1}{\beta_{c^m2}}[\omega_{c^m} - \alpha_{c^m} - \beta_{c^m 	1}\ln(z_{1})-a_{c^m}(y)-b_{c^m}(y^2) \\
	&- \ln(\textbf{w}')\bm{\lambda}_{c^m} - \textbf{e}'\bm{\delta}_{c^m} - \textbf{m}'\bm{\psi}_{c^m} - \bm{\tau}_{c^m}  - \varepsilon_{c^m}].
	\end{split}
	\end{equation}
	
Substituting equation (10) in $\tilde{\textbf{g}}(w^m, w^f,y,c^m,\textbf{z}_{-2};\bm{\xi})$, the demand for the remaining goods, we obtain the $z$-conditional demand system: 
	\begin{equation}
	\begin{split}
	\omega_{s} &= \varphi_s +  \theta_s\ln(z_{1}) + a_{s}(y)+b_{s}(y^2) + \frac{\beta_{s2}}{\beta_{{c^m}2}}\omega_{c^m}  \\
	-& \frac{\beta_{s2}}{\beta_{{c^m}2}}[a_{c^m}(y)+a_{c^m}(y^2) +\ln(\textbf{w}')\bm{\lambda}_{c^m}  + \textbf{e}'\bm{\delta}_{c^m} + \textbf{m}'\bm{\psi}_{c^m} + \bm{\tau}_{c^m}] + \zeta_s,
	\end{split}
	\end{equation}
where 
\vspace{2mm}
\begin{equation*}
	\begin{split}
		\varphi_s &= \alpha_s - \frac{\alpha_{c^m}\beta_{s2}}{\beta_{{c^m}2}}, \\
		\theta_s &= \beta_{s1}-\dfrac{\beta_{{c^m}1}\beta_{s2}}{\beta_{{c^m}2}}, \\
		\zeta_s &= \frac{\beta_{s2}}{\beta_{{c^m}2}}\varepsilon_{c^m} + \varepsilon_s
	\end{split}
\end{equation*}
for all goods $s \neq c^m$. One important source of endogeneity that arises from the estimation of (11), is the fact that the share of male private consumption is not independent of the new compound error term $\zeta_s$. A natural instrument for men's consumption is $z_2$ which satisfies the standard requirements for being a relevant and valid instrumental variable. It is worth noting that equation (10) demonstrates the correlation between $\omega_{c^m}$ and $z_2$, while the latter is excluded from equation (11). To mitigate this endogeneity problem, I employ a control function approach by incorporating the residuals from the first stage of the conditioning good into the estimation of equation (11).\footnote{Control functions for testing collective rationality are also used by \citeA{bobonis2009allocation, attanasio2014efficient, de2022household}.}

The exclusion restriction imposed by the collective model, as inferred from the estimation of the $z$-conditional demand system in equation (11), can be stated as follows:

\begin{equation}
	\frac{\partial \omega_s}{\partial \ln (z_{1})} =\theta_s  = 0 \quad\quad \forall \hspace{2mm} s \neq {c^m}.
\end{equation}
For each budget share equation in the system (11), a linear test is conducted to assess the significance of the parameter estimate of the relative personality factor. Restriction (12) indicates that once we condition the demand for the remaining goods on the demand for $c^m$, which is monotonically related to $z_2$, the additional variation provided by $z_1$ does not play a significant role in determining the household equilibrium. This condition is equivalent to the requirement of distribution factor proportionality, as discussed in \citeA{bourguignon2009efficient}. The exclusion restriction stated in equation (12) carries greater empirical power compared to the cross-equation restrictions presented in (9). This observation further strengthens the robustness of the estimation results obtained for the unconditional demand system.

\section{Empirical results}
I estimate the unconditional demand system in equation (8) using ordinary least squares (OLS), while the $z$-conditional demand system in equation (11) is estimated using a control function approach. In the control function approach, I incorporate the residuals obtained from a first-stage regression of male private consumption into the demand for the other commodities. To account for heteroskedasticity, I use robust standard errors, and I cluster the standard errors at the household level in all specifications. 

Table 3 presents the estimates of the system of unconditional demand equations. The specifications include several control variables: a linear control function for household full income and its square, instrumented with household potential income; the logarithm of spouses' wages and the interaction between them; the square of the husband's wage; husband's age and its square; husband's educational level; the number of children in the couple; a dummy variable indicating whether the couple is married or cohabiting; and time dummies.\footnote{Wife's age and educational level are not included in the specifications due to multicollinearity issues, as there is a significant positive assortative mating in age and education in the sample. However, the results remain robust when using the wife's characteristics as controls instead.} Additionally, I include the logarithm of the principal components in levels and their squares as additional taste shifters in the specifications. The marriage market personality ratios are also included in the analysis.

Firstly, it is observed that the \textit{relative} endowments of personality between spouses have a significant impact on male private consumption, leisure time, and public expenditures. Both personality factors positively affect private and public consumption, but negatively influence the allocation of leisure. Secondly, the second distribution factor, which is associated with conscientiousness and neuroticism, has a larger average effect compared to the first distribution factor. Thirdly, the ratios of the estimated coefficients of the distribution factors across commodities, as indicated in equation (9), are 0.37 for $c^m$, 0.17 for $c^f$, 0.16 for $l^m$, 0.37 for $l^f$, and 0.31 for $C$. These proportional average effects across commodities are supported by the results of the (bootstrapped) proportionality test presented at the bottom of Table 3. This evidence suggests that relative personality influences an individual's consumption within a partnership, but solely through its impact on the distribution of power within the household. Finally, personality also directly affects the allocation of resources through its impact on preferences, as evidenced by the significant effect of the principal components in \textit{levels} on commodities. Due to multicollinearity, the estimated coefficients of the wife's principal components are not included in the analysis.

\begin{table}[hbt!]
\caption*{Table 3. OLS estimates of the effect of relative personality on household consumption. System of unconditional demand functions. }
\centering

\begin{tabular}{lccccc}
\hline \hline
& \multicolumn{5}{c}{ \underline{Dependent variable: budget share}} \\
& $\omega_{c^{m}}$ & $\omega_{c^{f}}$ & $\omega_{\ell^{m}}$ & $\omega_{\ell^{f}}$ & $\omega_{C}$ \\ 
 \hline
\multirow{2}{*}{$\ln(\frac{PC1^f}{PC1^m})$}  & .033$^{\bm{+}}$ & .014   &-.022&-.082$^{\bm{+}}$ & .058\\
										     &(.016)&(.014)&(.024)&(.035)&(.044)\\
\multirow{2}{*}{$\ln(\frac{PC2^f}{PC2^m})$} & .088$^{\bm{+}}$ &.078&-.136$^{\bm{+}}$&-.218$^{\bm{+}}$& .186$^{\bm{+}}$\\
											&(.043)&(.054)&(.073)&(.097)&(.110)\\ \hline 
\multirow{2}{*}{$\ln(PC1^m)$}  & .036$^{\bm{+}}$  & .005 &-.037 &-.123$^{\bm{+}}$ &.119$^{\bm{+}}$\\													&	 (.019)& (.020)&(.031)&(.045) &(.056)\\
\multirow{2}{*}{$\ln(PC2^m)$}  & .112$^{\bm{+}}$ & .057 & -.231$^{\bm{+}}$& -.243&.305$^{\bm{+}}$\\											& (.068) & (.087) & (.105)&(.152)&(.188)		
\\
\hline 											
Additional covariates & Yes& Yes& Yes& Yes& Yes \\ 
Time dummies & Yes& Yes& Yes& Yes& Yes \\ \hline
Proportionality test & \multicolumn{5}{c}{$\chi^2$(4) = 0.892 ($p-$value =   .911)} \\ \hline \hline
\end{tabular}
\parbox{\textwidth}{\scriptsize%
\vspace{1eX} 
Notes: Sample size of 1130 couples. Robust standard errors clustered at the household level are in parentheses. $PC$: principal component. I estimate the proportionality test's $p-$value on its bootstrap distribution over 1000 replications. Additional covariates: linear control function for full income and its square instrumented with household potential income; the log of spouses' wages and the interaction between them; the square of husband's wage; husband's age and its square; husband's educational level; the number of children the couple has; marital status; the squares of the log of spouses' PCs in levels; and personality ratios. $^{\bm{+}}:$ Significant with at least 90\% of confidence.}
\end{table}

Table 4 presents the estimates of the $z$-conditional demand functions based on equation (11). The same control variables are used as in the unconditional demand equations. It should be noted that the conditioning good is $c^m$, and the relative level of PC2 is employed to invert the demand for this good. Importantly, both personality factors have a significant impact on the budget share equation of $c^m$. The most compelling evidence is obtained from estimations where the budget share equation is responsive to both factors \cite{de2022household}. Additionally, the relative level of PC2 is statistically significant in four out of five budget share equations. In the unconditional demand system (Table 3), the relative amount of PC1 is significant for two commodities (male private consumption and female leisure). However, in the $z$-conditional demand system (Table 4), it is not significant in any case. This evidence suggests that the impact of relative personality is indeed one-dimensional, meaning that relevant information regarding the intrahousehold allocation of resources is \textit{completely} summarized by the share of male private consumption. Crucially, this finding is confirmed by the result of the collective test at the bottom of Table 4.

\begin{table}[hbt!]
\caption*{Table 4. OLS estimates of the effect of relative personality on household consumption. System of $z$-conditional demand functions.}
\centering
\begin{tabular}{lcccc}
\hline \hline
 \multicolumn{5}{r}{\hspace{3cm} \underline{Dependent variable: budget share}} \\
&  $\omega_{c^{f}}$ & $\omega_{\ell^{m}}$ & $\omega_{\ell^{f}}$ & $\omega_{C}$  \\ \hline
\multirow{2}{*}{$\ln(\frac{PC1^f}{PC1^m})$} &    -.031 & .055  &.043 & -.067 \\
	             							&   (.034)& (.047) &(.064)&(.073) \\
\hline 
Additional covariates & Yes & Yes & Yes & Yes \\
Time dummies & Yes& Yes& Yes& Yes \\ \hline
Collective test & \multicolumn{4}{c}{$\chi^2$(4) =   3.476 ($p-$value =   .502)} \\ \hline \hline
\end{tabular}
\parbox{\textwidth}{\scriptsize%
\vspace{1eX} 
Notes: Sample size of 1130 couples. Robust standard errors clustered at the household level are in parentheses. $PC$: principal component. The conditioning good is $c^m$. I estimate the collective test's $p-$value on its bootstrap distribution over 1000 replications. Additional covariates: linear control function for full income and its square instrumented with household potential income; the log of spouses' wages and the interaction between them; the square of husband's wage; husband's age and its square; husband's educational level; the number of children the couple has; marital status; the square of the log of spouses' PCs in levels; and personality ratios. }
\end{table}

\pagebreak
\section{Personality and intrahousehold consumption inequality}

After providing theory-based evidence that (relative) personality affects the bargaining weights of household members, it is important to explore the relationship between personality and within-family inequality. Following the approach of \citeA{cherchye2020marital}, I analyze intrahousehold consumption inequality using the women and men relative individual cost of equivalent bundle (RICEB). For a given couple, these bundles are defined as follows:

\begin{equation}
	\begin{split}
		\textrm{RICEB}^i &= \frac{c^i + w^i\ell^i+C}{y} \quad\quad 	  \textrm{with} \hspace{2mm} i \in \{m,f\}.
	\end{split}
\end{equation}

Member-specific RICEBs describe how household members allocate consumption relative to the household's full income, taking into account both scale economies and the intrahousehold division of resources, thus providing an assessment of individual welfare.\footnote{It is worth noting that while the concept of RICEBs is related to the sharing rule concept in the collective literature, the RICEBs evaluate public expenditures at market prices instead of Lindahl prices. \citeA{bostyn2022time} utilize RICEBs to analyze individual welfare in a collective model that incorporates marriage market restrictions.} In this study, intrahousehold consumption inequality is proxied by the difference between partners' RICEBs, specifically RICEB$^f$ minus RICEB$^m$.

Next, I examine the distribution of intrahousehold consumption inequality for three categories of couples: (a) households where the female fraction of a specific personality trait is \textit{above} the 80th percentile of the distribution of all female fractions; (b) households where the female fraction of a specific personality trait is \textit{between} the 45th and 55th percentiles of the distribution of all female fractions; and (c) households where the female fraction of a specific personality trait is \textit{below} the 20th percentile of the distribution of all female fractions. This categorization allows for a comparison between households where the within-household female personality fraction is either high, moderate, or relatively low. It is important to note that the female personality fraction ($r_p$) for personality trait $p$ is constructed as $r_p = p^f/(p^f+p^m)$.\footnote{Appendix C provides a detailed overview of the distribution of these female personality fractions as well as the RICEB measures.} 

Figure 2 illustrates how intrahousehold consumption inequality varies with the \textit{relative} amount of personality within couples for each personality measure, comparing the three types of households mentioned above.  First, it can be observed that couples with a \textit{moderate} within-family difference in personality tend to exhibit, on average, a smaller degree of intrahousehold consumption inequality (indicated by the red dashed lines, which are more concentrated around zero on the horizontal axis). Second, for almost all personality measures (with the exceptions of openness and neuroticism), the black solid line is consistently positioned to the right of the blue dash-dotted line. This implies that a larger fraction of a woman's personality relative to her partner is associated with a greater allocation of intrahousehold resources towards her. This pattern is particularly pronounced for conscientiousness, self-esteem, and cognitive engagement.\footnote{Note that self-esteem and conscientiousness exhibited the highest loadings among the personality traits in the principal components analysis (see Table 2).} Indeed, in those three cases, as demonstrated in Panel A of Table 5, I strongly reject the null hypothesis of equal means between couples with a \textit{large} and \textit{small} female personality fraction (referring to the black and blue distributions in Figure 2). In Panel B of Table 5, I present the difference in average intrahousehold consumption inequality between households with \textit{large} and \textit{small} personality fractions in the sample. For instance, in couples where women exhibit higher levels of cognitive engagement than their male partners, there is an average of 4.48\% more intrahousehold resources allocated to them compared to couples where men are more engaged in intellectual activities.

\begin{table}[hbt!]
\caption*{Table 5. Panel A: Test of equal mean in intrahousehold inequality between couples with \textit{large} and \textit{small} female personality fractions. Panel B:  Difference in average intrahousehold inequality between couples with \textit{large} and \textit{small} female personality fractions.}
\centering
\begin{tabular}{l|cc|c}
\hline \hline
 &\multicolumn{2}{c}{\underline{\textbf{Panel A:}}} & \multicolumn{1}{c}{\underline{\textbf{Panel B:}}}\\
 &\multicolumn{2}{c}{\underline{\textbf{Bootstrap  statistics}}} &  \multicolumn{1}{c}{\underline{\textbf{Difference in inequality}}} \\
  &    $t$-statistic & $p$-value \\ \hline 
Agreeableness & -.411 &.468 & 0.428\% \\
Openness &  -1.609 &.225&1.851\% \\
Extraversion &  -1.097 &.349&1.213\% \\
Conscientiousness &  -3.506$^{\bm{+}}$ &.014& 3.949\%\\
Neuroticism & .400&.484&0.476\%\\
Self-esteem &  -4.022$^{\bm{+}}$ & .005 & 3.441\%\\
Cognitive engagement &  -3.776$^{\bm{+}}$&.009 & 4.486\%
\\ \hline \hline
\end{tabular}
\parbox{\textwidth}{\scriptsize%
\vspace{1eX} 
Notes: Panel A shows the results of a bootstrapped $t$-test of equal mean between the black and blue distributions shown in Figure 2. I estimate both the $t$-statistic and $p$-value on their bootstrap distribution over 1000 replications. Panel B shows the difference in the average intrahousehold inequality between black and blue distributions shown in Figure 2. \\ $^{\bm{+}}$ Significant with at least 90\% of confidence.}
\end{table}

\begin{figure}[ht]
\rotatebox[origin=l]{90}{Figure 2. Intrahousehold consumption inequality and relative personality. \strut} \hspace{0.5cm}
    \includegraphics[scale=.6, angle=90]{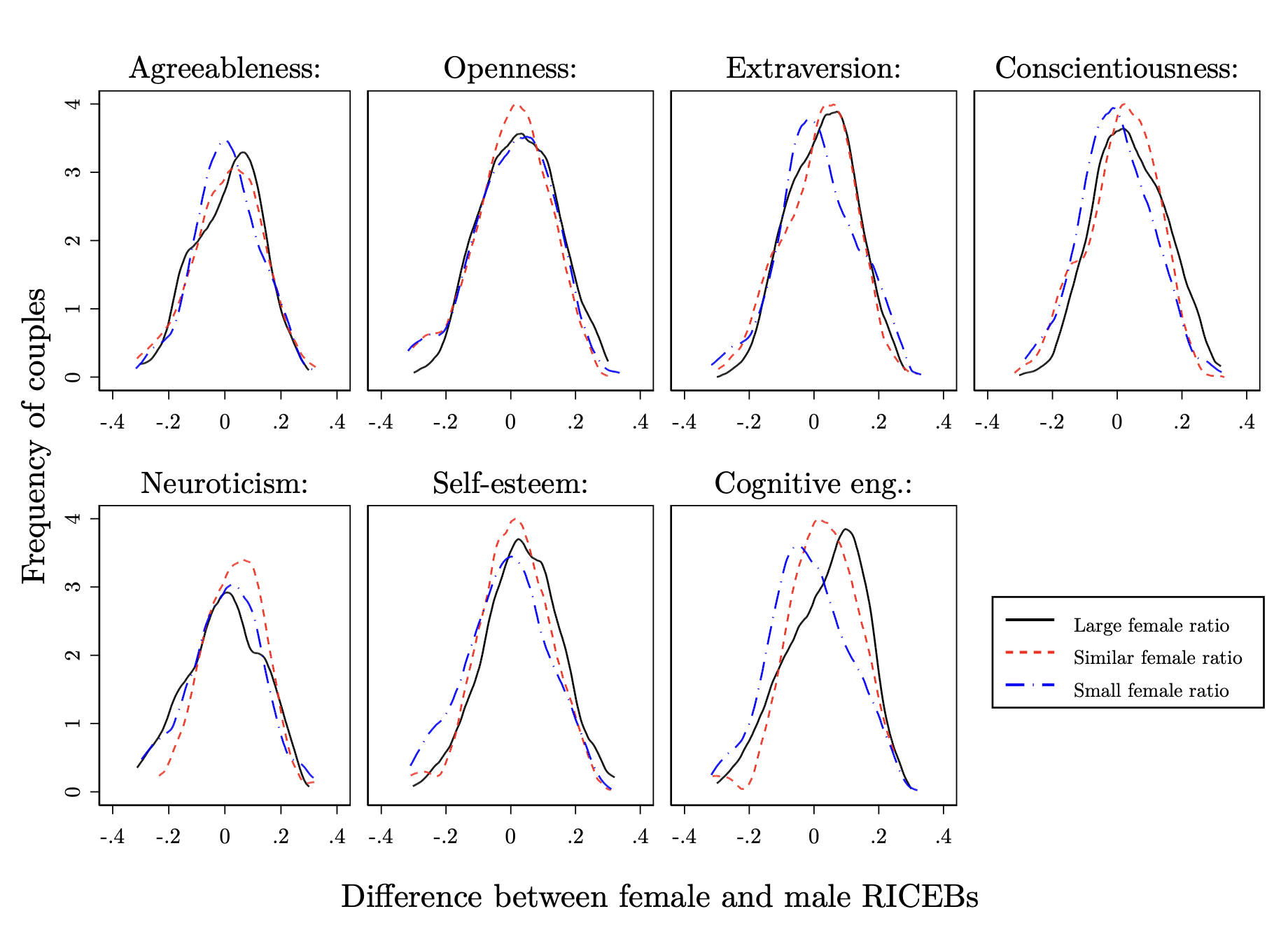}
    \label{fig:PropProf}
   \rotatebox{90}{\footnotemark \hspace{0.1cm}  { {\scriptsize Notes: This figure shows kernel density plots of intrahousehold inequality (i.e., RICEB$^f$ minus RICEB$^m$) and within-couple female personalities ratios.  }} } 

 \end{figure}

\section{Conclusion}

This paper presents compelling evidence, based on theoretical foundations, regarding the role of personality in resource allocation within households when assuming Pareto-efficient decision-making. By examining variations in personality traits among Dutch couples, this study tests for distribution factor proportionality and the exclusion restriction utilizing a conditional demand system estimation. The findings do not allow for the rejection of the hypothesis that (relative) personality influences the bargaining process within households. Notably, women who exhibit higher levels of conscientiousness, self-esteem, and cognitive engagement relative to their male partners tend to receive a larger proportion of intrafamily resources. To address potential selection bias in personality, the budget share equations are conditioned on the level of personality and additional explanatory variables that capture the structure of the marriage market in relation to personality traits within the sample.

The findings presented in this paper provide strong support for conducting a more comprehensive and structural analysis to explore the significance of personality traits within the family context, as well as the underlying mechanisms through which these traits exert their influence. Firstly, employing a model with a more robust parametric structure for preferences and the sharing rule, similar to approaches utilized by \citeA{browning2013estimating} or \citeA{cherchye2017household}, would offer deeper insights into the welfare implications of personality traits. Such an approach could enhance our understanding of how these traits affect individual well-being. Secondly, it is worth noting that several studies have demonstrated the importance of personality traits in assortative mating within the marriage market (\citeA{lundberg2012personality} or \citeA{dupuy2014personality}). Therefore, it would be valuable to estimate a matching model and examine the complete structure of the marriage market as a potential driver of power dynamics. This would allow for a comprehensive assessment of how personality traits shape partner selection and subsequent resource allocation within households. Lastly, the current paper's framework overlooks intertemporal aspects that are relevant to household consumption, such as the influence of personality on occupational or educational choices (\citeA{todd2020dynamic}). Considering these factors in future research would enhance the richness and applicability of the analysis.

\FloatBarrier   

\newpage
\section*{Appendix}
\appendix
\section{Stability of personality traits}
This section illustrates the evolution of personality over time for women and men in our sample. Figures A1 and A2 show the average score by age for each personality measure. I consider all waves together. 

\begin{figure}[hbt!]
\caption*{Figure A1. Female average personality scores by age.}
\centering
\includegraphics[scale=0.5]{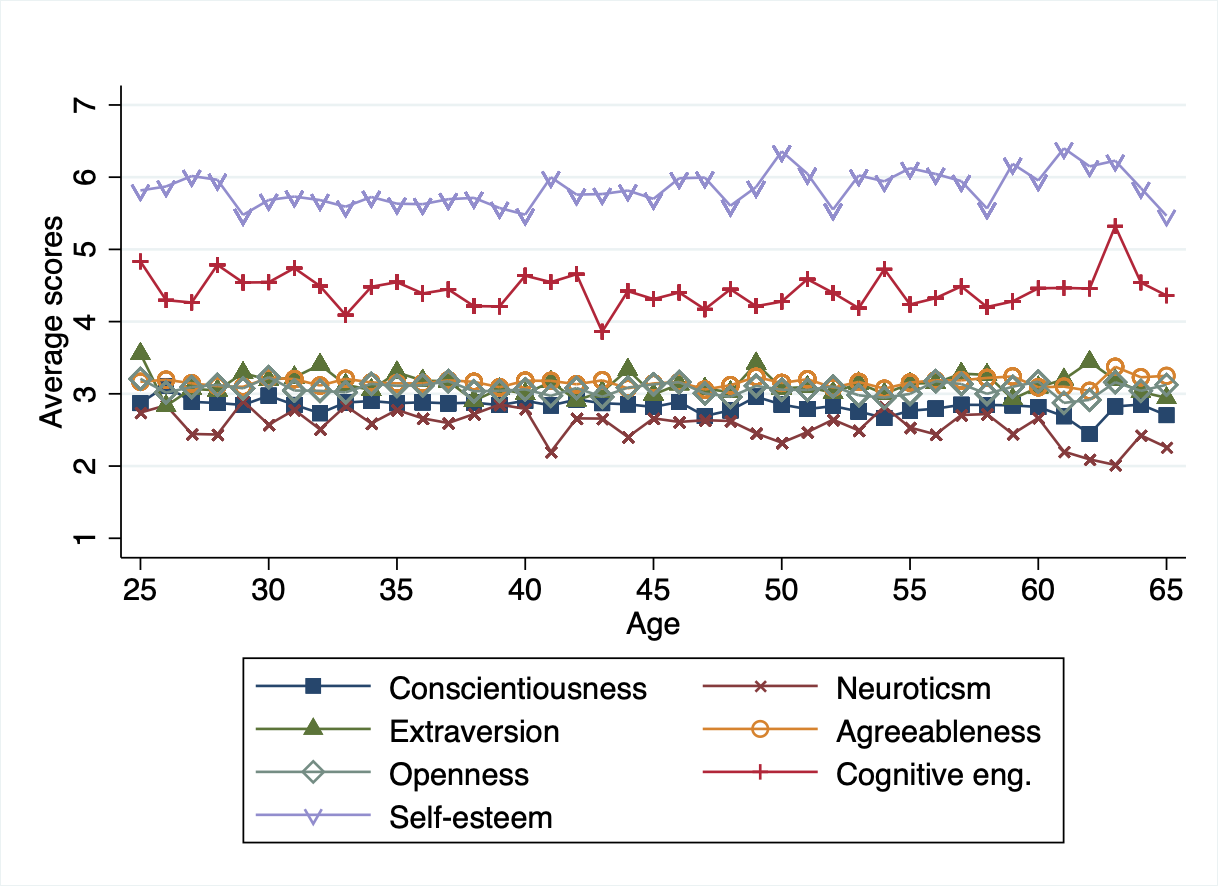}
\end{figure}

\begin{figure}[hbt!]
\caption*{Figure A2. Male average personality scores by age.}
\centering
\includegraphics[scale=0.5]{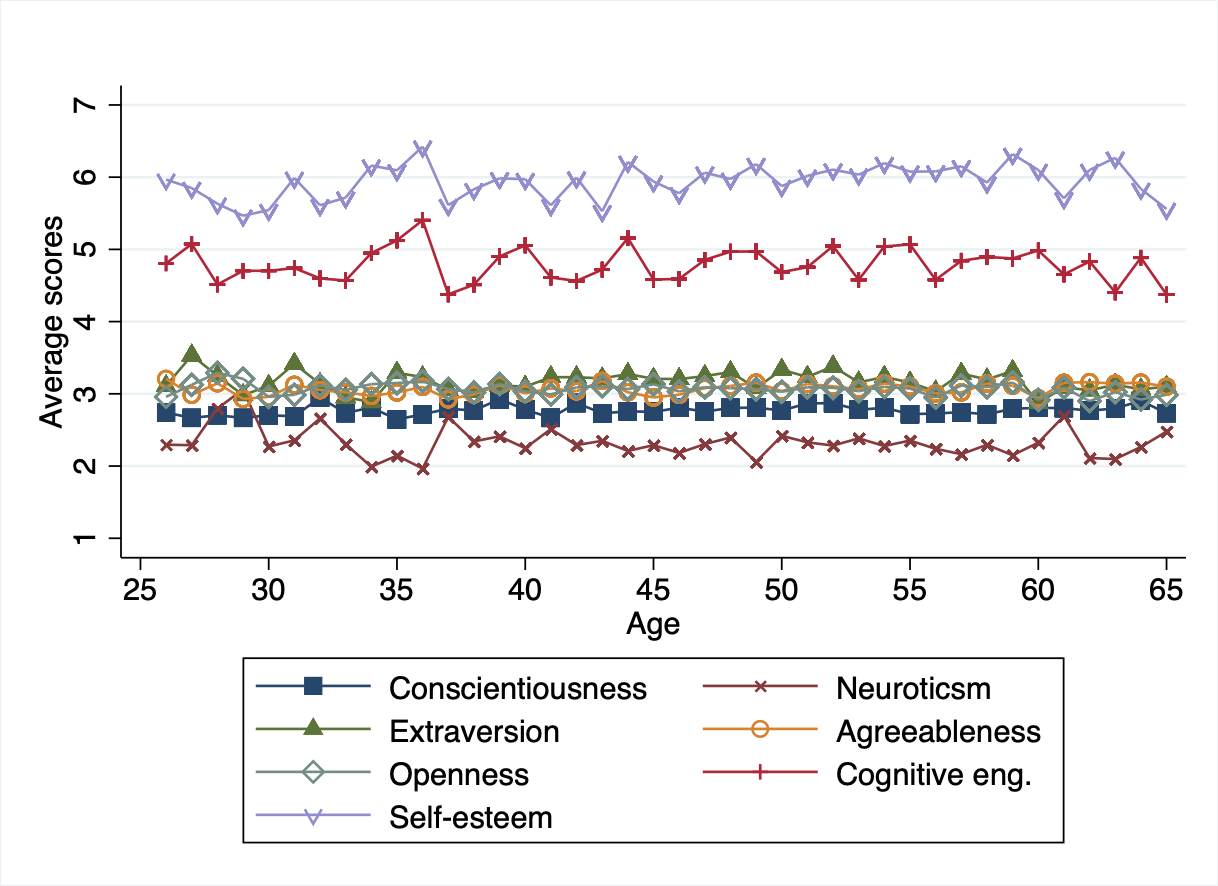}
\end{figure}

\FloatBarrier

\section{Monotonic relationship between $z_2$ and male private consumption ($\omega_{c^{m}}$)}

%

Following \citeA{attanasio2014efficient}, I study the relationship between the second distribution factor ($z_2 = \ln(\frac{PC2^f}{PC2^m})$) and the share of male private consumption ($\omega_{c^{m}}$) by looking at the point estimates of different polynomials. The direction of the point estimates implies an increasing relationship between the share of men's private expenditures and the second measure of relative personality within households. This information, together with the fact that both distribution factor influence significantly $\omega_{c^{m}}$ (see Table 3), support the choice of men's private expenditures as the conditioning good.

\begin{table}[hbt!]
\caption*{Table A1. Effect of distributions on the consumption shares.}
\centering

\begin{tabular}{lccccc}
\hline \hline
& \multicolumn{5}{c}{ \underline{Dependent variable: budget share}} \\
& $\omega_{c^{m}}$ & $\omega_{c^{f}}$ & $\omega_{\ell^{m}}$ & $\omega_{\ell^{f}}$ & $\omega_{C}$ \\ 
 \hline
\multirow{2}{*}{$\ln(\frac{PC1^f}{PC1^m})$}  & .022 & .033$^{\bm{+}}$&-.019& -.104$^{\bm{+}}$ & .068\\
										     &(.021)&(.017)&(.029)&(.039)&(.052)\\
\multirow{2}{*}{$[\ln(\frac{PC1^f}{PC1^m})]^2$}  &  -.001 & .005$^{\bm{+}}$   &-.001&-.004 & .002\\
										     &(.001)&(.001)&(.002)&(.003)&(.005)\\
\multirow{2}{*}{$[\ln(\frac{PC1^f}{PC1^m})]^3$}  & .0002& .001   &-.003&.001 & .0002\\
										     &(.001)&(.001)&(.002)&(.003)&(.003)\\	\hline		     
\multirow{2}{*}{$\ln(\frac{PC2^f}{PC2^m})$} & .010 &.029& -.158$^{\bm{+}}$& -.106& .225$^{\bm{+}}$\\
											&(.051)&(.072)&(.084)&(.115)&(.139)\\
\multirow{2}{*}{$[\ln(\frac{PC2^f}{PC2^m})]^2$} & .002 & -.005&-.006&-.0004& .010\\
											&(.004)&(.005)&(.007)&(.013)&(.013)\\	
											
\multirow{2}{*}{$[\ln(\frac{PC2^f}{PC2^m})]^3$} & .014$^{\bm{+}}$ &.002&-.002&-.018& .004\\
											&(.005)&(.005)&(.007)&(.014)&(.015)\\										
\hline 
\hline 											
Additional covariates & Yes& Yes& Yes& Yes& Yes \\ 
Time dummies & Yes& Yes& Yes& Yes& Yes \\ \hline \hline 
\end{tabular}
\parbox{\textwidth}{\scriptsize%
\vspace{1eX} 
Notes: Sample size of 1130 couples. Robust standard errors clustered at the household level are in parentheses. $PC$: principal component. Additional covariates: linear control function for full income and its square instrumented with household potential income; the log of spouses' wages and the interaction between them; the square of husband's wage; husband's age and its square; husband's educational level; the number of children the couple has; marital status; the log of spouses' PCs in levels and their squares; and personality ratios. $^{\bm{+}}:$ Significant with at least 90\% of confidence.}
\end{table}

\pagebreak
\section{Distribution of female personality fractions and RICEBs}
\FloatBarrier
\begin{table}[hbt]
\caption*{Table C1. Summary statistics for female personality ratios and RICEBs measures ($N$ = 1130 couples)}
\centering
\begin{tabular}{lccccccc}

\hline \hline
& Mean & Std. Dev. & Min & p25 & Median & p75 & Max\\  \hline
RICEB$^f$ &  0.606   &  0.070&     0.273&     0.563&     0.609     &0.653     &0.802 \\
RICEB$^m$	&  0.598  &   0.072  &   0.323&     0.551    & 0.595&     0.644&     0.839 \\ \hline
Female fractions ($r_p$): \\
\hspace{2mm}\textit{Self-Esteem}&  0.494   &  0.037    & 0.391&     0.474     &0.496     &0.515     &0.631\\
\hspace{2mm}\textit{Extraversion}&  0.495   &  0.062  &   0.286&     0.457     &0.488&     0.535     &0.700\\
\hspace{2mm}\textit{Openness} & 0.500   &  0.031     &0.357     &0.480  &   0.500     &0.520 &    0.694\\
\hspace{2mm}\textit{Neuroticism}&0.531   &  0.076     &0.309     &0.475 &    0.532     &0.580  &   0.773 \\
\hspace{2mm}\textit{Agreeableness}&  0.508&     0.027     &0.413  &   0.490     &0.509   &  0.521 &    0.636\\
\hspace{2mm}\textit{Conscientiousness}&0.507 &    0.033    & 0.356 &    0.489     &0.505    & 0.529&     0.612 \\
\hspace{2mm}\textit{Cognitive Engagement}& 0.479&     0.064&     0.299&     0.435&     0.478&     0.518    & 0.679\\

\hline 									\hline 								
\end{tabular}
\end{table}

\newpage
\bibliographystyle{apacite}
\bibliography{Bib.bib}
\end{document}